\documentclass[aps,pre,twocolumn,superscriptaddress]{revtex4-1}
\usepackage{empheq}
\usepackage{amsmath}
\usepackage{tabularx}
\usepackage{dsfont}
\usepackage{graphicx}
\usepackage[hidelinks]{hyperref}
\usepackage[normalem]{ulem}
\usepackage{stackrel}
\usepackage{amssymb,latexsym,mathrsfs}
\usepackage{amsfonts}
\usepackage{srcltx}
\usepackage{color}
\usepackage{cancel}
\usepackage{stix}
\usepackage{xr}
\usepackage{placeins}
\usepackage{algorithm}
\usepackage{algpseudocode}
\usepackage{etoolbox}
\usepackage{multirow}

\newcommand{\la}{\langle}
\newcommand{\ra}{\rangle}

\newcommand*{\white}{\textcolor{white}}

\newcommand*{\Vghost}{\white{$a^{M^{M}}$}}
\newcommand*{\Vghostlarge}{\white{$a^{M^{M^{M}}}$}}

\newcolumntype{Y}{>{\centering\arraybackslash}X}
\newcolumntype{L}[1]{>{\raggedright\let\newline\\\arraybackslash\hspace{0pt}}m{#1}}
\newcolumntype{C}[1]{>{\centering\let\newline\\\arraybackslash\hspace{0pt}}m{#1}}
\newcolumntype{R}[1]{>{\raggedleft\let\newline\\\arraybackslash\hspace{0pt}}m{#1}}

\newcommand\CONDITION[2]%
{\begin{tabular}[t]{@{}l@{}l@{}}
		#1&#2
	\end{tabular}%
}
\algdef{SE}[WHILE]{While}{EndWhile}[1]%
{\algorithmicwhile\ \CONDITION{#1}{\ \algorithmicdo}}%
{\algorithmicend\ \algorithmicwhile}
\algdef{SE}[FOR]{For}{EndFor}[1]%
{\algorithmicfor\ \CONDITION{#1}{\ \algorithmicdo}}%
{\algorithmicend\ \algorithmicfor}
\algdef{S}[FOR]{ForAll}[1]%
{\algorithmicforall\ \CONDITION{#1}{\ \algorithmicdo}}
\algdef{SE}[REPEAT]{Repeat}{Until}{\algorithmicrepeat}[1]%
{\algorithmicuntil\ \CONDITION{#1}{}}
\algdef{SE}[IF]{If}{EndIf}[1]%
{\algorithmicif\ \CONDITION{#1}{\ \algorithmicthen}}%
{\algorithmicend\ \algorithmicif}%
\algdef{C}[IF]{IF}{ElsIf}[1]%
{\algorithmicelse\ \algorithmicif\ \CONDITION{#1}{\ \algorithmicthen}}

\makeatletter
\newcounter{phase}[algorithm]
\newlength{\phaserulewidth}
\newcommand{\setphaserulewidth}{\setlength{\phaserulewidth}}
\newcommand{\phase}[1]{%
	\vspace{-1.25ex}
	\Statex\leavevmode\llap{\rule{\dimexpr\labelwidth+\labelsep}{\phaserulewidth}}\rule{\columnwidth-9pt}{\phaserulewidth}
	\Statex\strut\refstepcounter{phase}\textbf{Phase~\thephase~--~#1}
	\vspace{-1.25ex}\Statex\leavevmode\llap{\rule{\dimexpr\labelwidth+\labelsep}{\phaserulewidth}}\rule{\linewidth-9pt}{\phaserulewidth}}
\makeatother
\setphaserulewidth{.7pt}

\begin{document}

\title{Fast algorithm for topologically disordered lattices with constant coordination number}
\author{Manuel Schrauth}
\author{Jefferson S. E. Portela}
\affiliation{Institute of Theoretical Physics and Astrophysics,	University of W\"urzburg, 97074 W\"urzburg, Germany}

\begin{abstract}
	We present a stochastic algorithm for constructing a topologically disordered (i.e., non-regular) spatial lattice with nodes of constant coordination number, the \emph{CC lattice}. The construction procedure dramatically improves on an earlier proposal [Phys.~Rev.~E.~\textbf{97}, 022144 (2018)] with respect to both computational complexity and finite-size scaling properties -- making the CC lattice an alternative to proximity graphs which, especially in higher dimensions, is significantly faster to build. Among other applications, physical systems such as certain amorphous materials with low concentration of coordination defects are an important example of disordered, constant-coordination lattices in nature. As a concrete application, we characterize the criticality of the 3D Ising model on the CC lattice. We find that its phase transition belongs to the clean Ising universality class, establishing that the disorder present in the CC lattice is a non-relevant perturbation in the sense of renormalization group theory.
\end{abstract}

\maketitle

\section{Introduction}
\label{sec:Introduction}

Given a set of discrete points in space, if connections between these points are created according to a rule defined in terms of geometrical closeness, one obtains a so-called \emph{proximity graph}. Such graphs possess only local connections and their typical shortest path length scales as $\bar{l}\sim N^{1/d}$ on a $d$-dimensional set of $N$ points, in contrast to small-world networks~\cite{watts1998} and some scale-free networks~\cite{cohen2003}, where shortcuts provided by long-range connections lead to scalings $\bar{l}\sim\ln N$ and $\bar{l}\sim\ln\ln N$, respectively.

The most prominent proximity graph in two dimensions is arguably the Delaunay triangulation (DT)~\cite{toth2017}. Its construction rule specifies that the circumcircle of any DT triangle must be empty, i.e., it cannot contain any points of the set. Furthermore, in the DT, it is guaranteed that the distance along the edges between any two points is not larger than about 2.42 times their metric distance~\cite{okabe2000}. Similar, but more restrictive rules lead to the relative neighborhood graph (RNG)~\cite{toussaint1980} and Gabriel graph (GG)~\cite{gabriel1969}, which can be constructed from the DT by removing specific bonds~\cite{jaromczyk1992}. Also the (first) nearest-neighbor graph turns out to be a subgraph of the DT. A further simple proximity construction is the random geometric graph (RGG)~\cite{barthelemy2011}, where any two points whose distance falls below a certain threshold are linked.

Proximity graphs are useful in a wide range of applications, most notably mesh generation, surface modeling, pattern classification, ad-hoc networks, path planning and astrophysics~\cite{aurenhammer1991,jaromczyk1992,weatherill1992,bebis1999,gopi2000,ramella2001,wu2007,barthelemy2011}. Besides topics predominantly related to computer science, proximity graphs constructed from random point sets have also found application in  statistical physics~\cite{chandler1987}, with focus on the behavior of critical phenomena~\cite{tauber2014} for systems placed on such a graph rather than on a regular lattice. These irregular lattices are said to present \emph{topological} disorder, in order to distinguish their disorder from that of, e.g., diluted regular lattice models (a research topic with an even longer history, see Ref.~\cite{martins2007} and references therein). In the language of renormalization group theory~\cite{pelissetto2002} the non-regular structure of those lattices represents a source of disorder, and the central question is whether it is \emph{relevant}, i.e., whether it changes the character of the phase transition of a given physical model. 

\begin{figure}[b]
	\centering
	\begin{minipage}{.5\columnwidth}
		\centering
		\includegraphics[width=\linewidth]{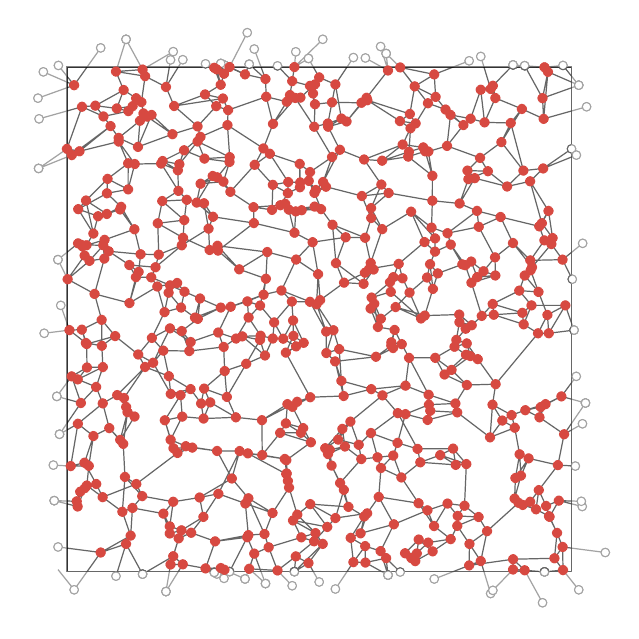}\\\vspace*{-2mm}(a)
	\end{minipage}%
	\begin{minipage}{0.5\columnwidth}
		\centering
		\includegraphics[width=\linewidth]{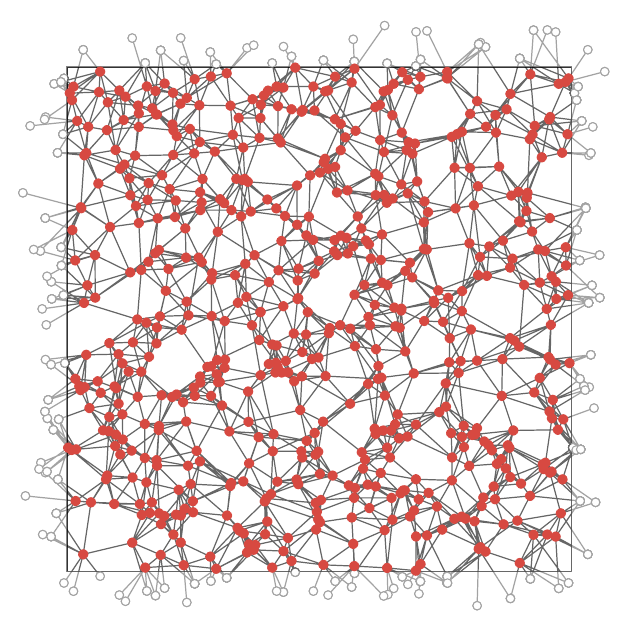}\\\vspace*{-2mm}(b)
	\end{minipage}
	\caption{Two-dimensional realizations of the CC lattice for coordination numbers (a) $q=4$ and (b) $q=8$ on sets of $N=24^2$ points with periodic boundaries.}
	\label{fig:cc_example}		
\end{figure}

\begin{figure*}[tb]
	
	\centering
	\begin{minipage}{.33\textwidth}
		\centering
		\includegraphics[width=\linewidth]{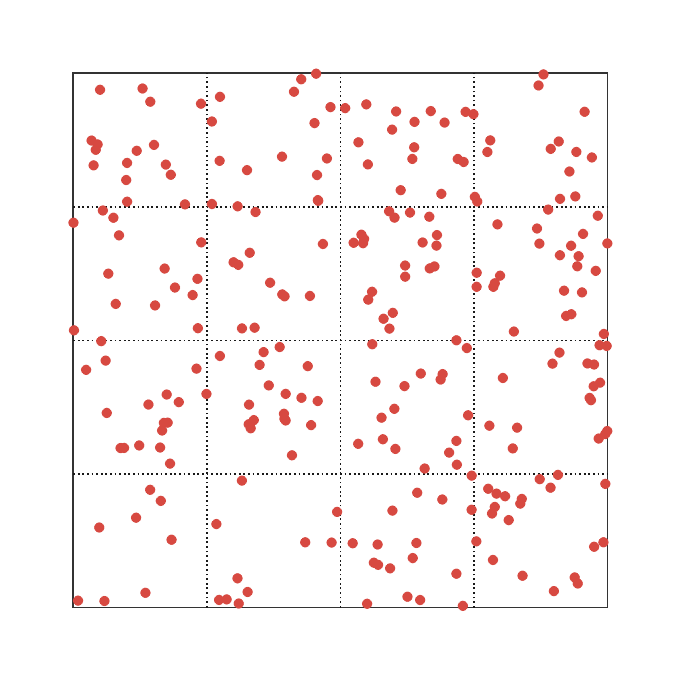} \\[-3mm](a)
		\includegraphics[width=\linewidth]{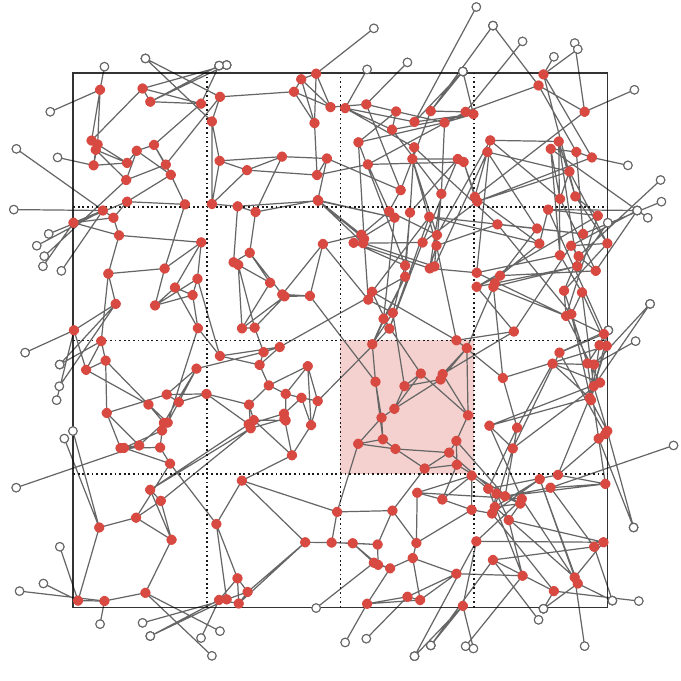} \\[-2mm](d)
	\end{minipage}
	\begin{minipage}{.33\textwidth}
		\centering
		\includegraphics[width=\linewidth]{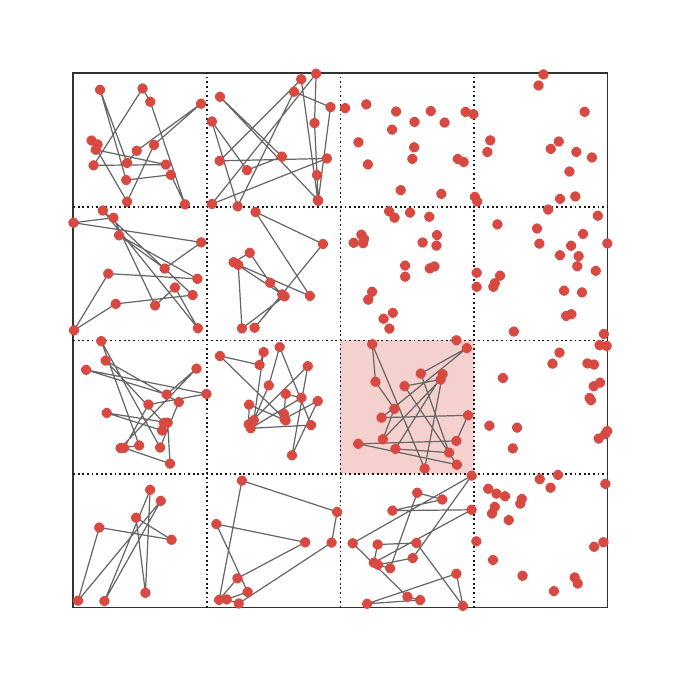} \\[-3mm](b)
		\includegraphics[width=\linewidth]{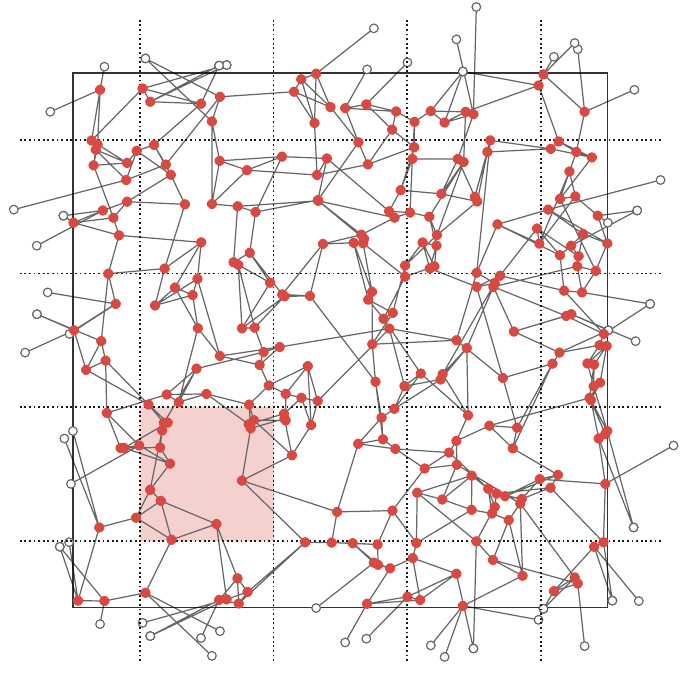} \\[-2mm](e)

	\end{minipage}
	\begin{minipage}{.33\textwidth}
		\centering
		\includegraphics[width=\linewidth]{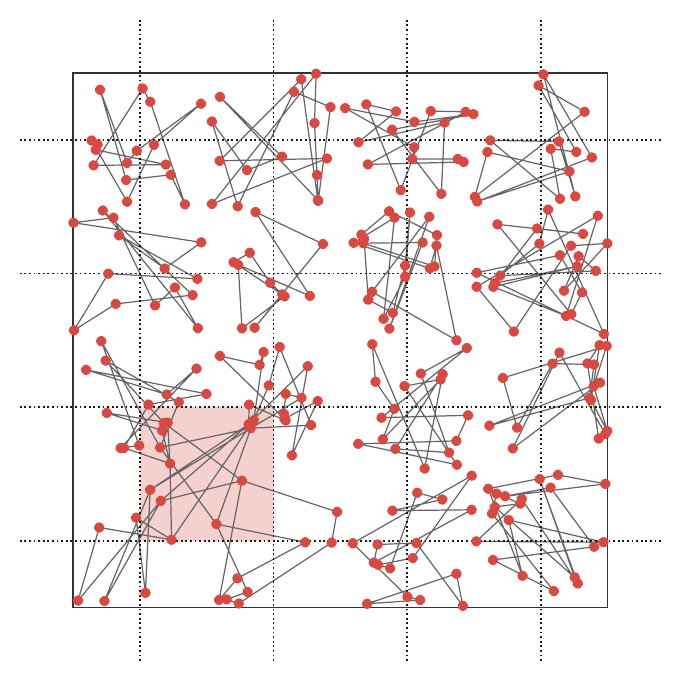} \\[-3mm](c)
		\includegraphics[width=\linewidth]{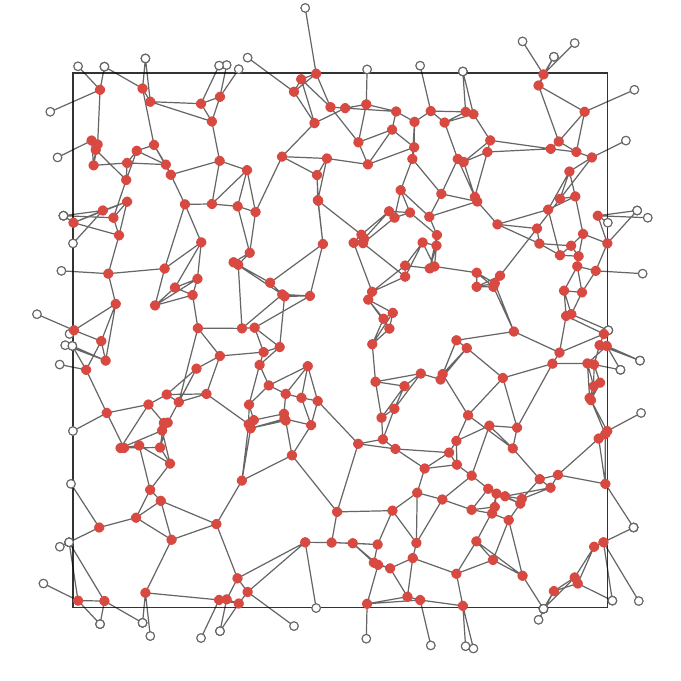} \\[-2mm](f)
	\end{minipage}
	\caption{Illustration of the construction process of a CC lattice with coordination number $q=4$: (a) sites are grouped into cells and (b) each site is randomly connected to $q_1=2$ neighbors in its respective cell; the sites are then (c) regrouped into staggered cells and each site is connected to $q_2=2$ additional random neighbors in its cell; bond lengths are finally minimized by simulated annealing dynamics, performed first (d) in the original cell partition and then (e)~for the staggered partition, resulting in (f) a final lattice with coordination number $q=q_1+q_2$. Note that the shaded region marks the currently processed cell in the respective construction step. 
	A further illustration of the construction is available as a video in the Supplementary Material.}
	\label{fig:cc_construction}		
\end{figure*}

Paradigmatic systems exhibiting continuous phase transitions include the equilibrium Ising model~\cite{ising1925} and the non-equilibrium contact process (CP)~\cite{hinrichsen2000}, both already studied on two-dimensional Delaunay triangulations using large-scale numerical Monte Carlo simulations~\cite{janke1993,janke1994,lima2000,janke2002,oliveira2008,oliveira2016}. For the CP, it has been shown that the topological disorder of the DT presents no relevant perturbation, and its phase transition remains in the universality class of directed percolation. This result cannot be explained by Harris' seminal relevance criterion~\cite{harris1974,harris2016}, and it motivated the introduction of a generalization~\cite{barghathi2014} that attributes the non-relevance to strong spatial anti-correlations in the coordination numbers of the lattice nodes. Also for this new criterion, however,  violations have been found~\cite{schrauth2018b}, meaning that a complete criterion explaining the influence of topological disorder on continuous phase transitions still remains to be found.

Fluctuating local coordination numbers are typical for proximity graphs (DT, RNG, GG, RGG, $k$-nearest neighbor, \ldots). Therefore, a lattice which is still disordered in the topological sense, but where coordination number fluctuations are absent can be an interesting tool for investigating the influence of disorder on phase transitions in physical systems. One such lattice is the Constant Coordination (CC) lattice, which we introduced with J.~A.~J.~Richter in~\cite{schrauth2018a}, and is illustrated in Fig.~\ref{fig:cc_example}. By construction, every site in the CC lattice is connected to exactly $q$ other sites, without allowing for self or double connections. In this way, any perturbation on the phase transition of a model can only originate from the implicit connectivity disorder, independent from coordination numbers.

In this paper we refine the original construction rules~\cite{schrauth2018a} and present an improved local algorithm with significantly reduced computational complexity (from quadratic to linear in the number of points), which is straightforward to apply to dimensions larger than two and also eliminates a few shortcomings related to the applicability to finite-size scaling~\cite{cardy2012} simulations. Furthermore, we make use of the new algorithm to investigate the 3D Ising phase transition on the CC lattice -- an undertaking unfeasible with the original algorithm.

The paper is organized as follows. In Sec.~\ref{sec:BasicConcept} we illustrate the basic idea of the construction procedure. In Sec.~\ref{sec:AlgorithmicDetails} we review the drawbacks of the originally proposed algorithm and show how they are resolved in the current approach. After a few remarks on the immediate generalization to higher dimensions in Sec.~\ref{sec:HigherDimensions}, we present and analyze, as an example of application, numerical Monte-Carlo simulations of the Ising model on a 3D CC lattice with coordination number $q=4$ in Sec.~\ref{sec:Application}. Finally, in Sec.~\ref{sec:Conclusion} we present our concluding remarks.

\section{Basic Concept}
\label{sec:BasicConcept}

\begin{figure}[t]
	\centering
	\includegraphics[width=\linewidth]{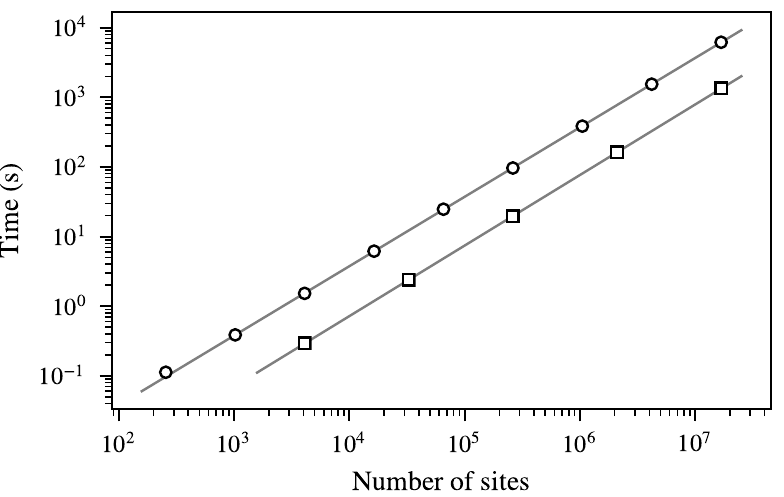}
	\caption{Computing time for CC lattices with $q=4$ for sizes $L=16,32,\ldots,4096$, in 2D (circles), and $L=16,32,\ldots,256$, in 3D (squares). The straight lines are fits to the data points that return exponents 0.997 and 1.013, supporting the linear scaling behavior. All data points are averages from at least ten realizations on an Intel Core i7-6700 CPU. The error bars are smaller than the marker size.}
	\label{fig:scaling}		
\end{figure}

As typical for topologically disordered random lattices, our starting point is a cloud of randomly distributed points in a toroidal domain of linear size $L$ with Euclidean metrics. In the first step of the lattice construction, bonds between random pairs of sites are gradually introduced until each site has exactly $ q $ neighbors. The key step after that, is to subject the graph to a dynamical rewiring, by means of a simulated annealing (SA) procedure \cite{kirkpatrick1983}, in order to achieve locality, i.e., to keep connections effectively short ranged. Specifically, in the SA algorithm, two bonds $\overline{il}$ and $\overline{jk}$ are taken at random and rewired to a new configuration $\overline{ij}$ and $\overline{kl}$ whenever this leads to a decrease of the sum of the bond lengths, i.e., whenever
\begin{align}
d(i,j)+d(k,l) < d(i,l) +d(j,k).\label{eq:constraint}
\end{align}
If, instead, the new configuration leads to an increase of the combined link lengths, the rewiring is accepted only with probability $ \exp(-\Delta H/T) $, where
\begin{align}
\Delta H \equiv [d(i,j)+d(k,l)] - [d(i,l)+d(j,k)] \label{eq:cost}
\end{align}
defines the cost function. The simulated annealing temperature $T$ has the effect of noise on the convergence to a state of low cost function. The value of $T$ is logarithmically decreased during the simulation, such that in the beginning, non-optimal rewiring updates are accepted with moderate probability, whereas in the final stages, this probability almost vanishes and only those moves are performed where condition~\eqref{eq:constraint} strictly applies.

The first algorithm for the CC construction, put forward in the original proposal~\cite{schrauth2018a}, presents two central drawbacks: first, it is computationally expensive; second, it requires considerable care and the introduction of an inconvenient extra parameter in order to avoid any dependence of its geometrical characteristics on the lattice size.

An improved algorithm for generating the CC lattice, which overcomes these drawbacks, can be obtained based on a simple key concept: instead of over the whole lattice, we perform the construction locally, in subgraphs delimited by grid cells of a small, fixed size. That brings the complexity from $\mathcal{O}(N^2)$ to $\mathcal{O}(N)$ by keeping fixed the size of the set to which simulated annealing is applied, and its locality also guarantees there are no lattice-size dependencies.

\section{Algorithmic Details}
\label{sec:AlgorithmicDetails}

\begin{figure}[tb]
	\centering
	\includegraphics[width=\linewidth]{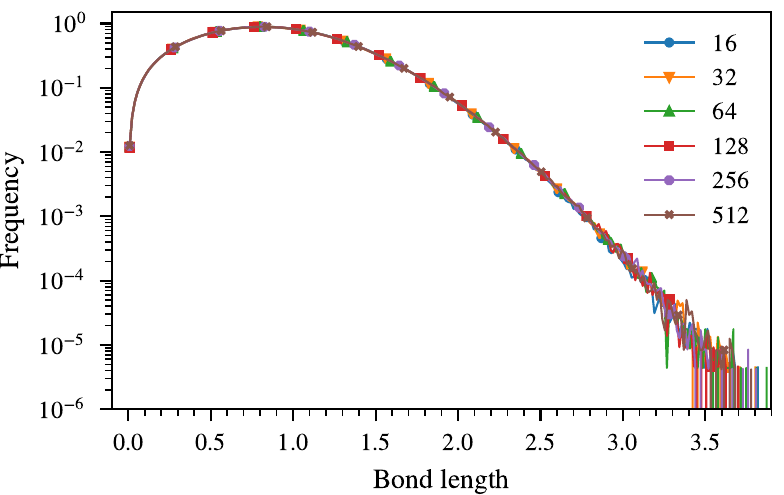}
	\caption{Bond length histograms of for two-dimensional CC lattices with $q=4$ for several values of $L$, indicating that the high degree of locality is independent of the lattice size. Each curve is computed from a sample of $10^8$ bonds, for typical parameter values.}
	\label{fig:histogram}		
\end{figure}

In this section we elaborate on the Constant Coordination lattice algorithm sketched in Sec.~\ref{sec:BasicConcept}, giving special attention to its improvements over the original proposal~\cite{schrauth2018a}, which presented important drawbacks:

\paragraph{Computational cost}

	Since every rewiring step takes, at random, two bonds of the full set, the time complexity scales as $ \mathcal{O}(sqN^2) $, where $N$ is the number of sites in the lattice, $q$ the coordination number and $s$ the number of SA steps. Even though $q$ and $s$ are constant parameters, the $ \mathcal{O}(N^2) $ dependence alone renders the algorithm prohibitively expensive for large lattices. Besides, as the typical bond length becomes small with respect to $L$, random rewirings grow increasingly unlikely to satisfy condition~\eqref{eq:constraint} and the majority of update attempts is rejected, resulting in a slow convergence which demands a very large number of steps~$s$.

\paragraph{Initial configuration} 

	The need for a large number of rewiring steps can be mitigated by starting the SA procedure from a configuration that already has a certain degree of locality, instead of being fully random. Such an initial configuration can be obtained simply by, considering only sites with fewer than $q$ bonds, randomly linking those sites to their nearest neighbors until each site has exactly $q$ bonds. This requires the full distance matrix of the sites to be known, which can be calculated in $\mathcal{O}(N\ln N)$ using spatial tree techniques~\cite{bentley1975}. In practice, this initial step allows the parameter $s$ to be reduced by about two orders of magnitude.

\paragraph{Pathological motifs}

	However, the use of an optimized initial configuration also comes with a drawback of its own: the occasional failure to produce a so-called simple graph, due to unlucky configurations, or \emph{pathological} motifs. One such configuration arises, e.g.,~when the last site in the initial construction loop is left to connect to itself (all the other sites already having $q$ bonds). A number of examples of such motifs can be found in Appendix~\ref{sec:PathologicalMotifs}. These occasional failures must be dealt with using either some involved iterative procedure or a complete restart.

\paragraph{Micro-scale equivalence} 

	A lattice with $N$ sites and an arbitrary subgraph with $N$ sites from a larger lattice should be indistinguishable with respect to their connectivity properties (such as average bond length, shortest path, clustering, etc.) up to boundary effects. This property is crucial for the application of finite-size scaling methods. While it is trivially fulfilled for any geometrically constructed lattice, like regular lattices, DT, RGG, etc., this is not the case in the original CC algorithm. There, even though the initial connection step connects most of the sites locally, typically a few larger bonds spanning a significant part of the system can not be avoided, thus introducing a dependence on the lattice size. Therefore, in order to achieve micro-scale equivalence, larger lattices must be subjected to longer SA procedures, i.e., instead of a constant parameter $s$, we have a function $s=s(L)$, which is an undesirable additional parameter and a possible source of error.

\begin{figure}
	\hrule height \phaserulewidth
	\begin{algorithmic}
		\State \textbf{Input}: set $G$ of $N=L^d$ sites in $\left[0,L\right)^d$\Vghostlarge
		\State \textbf{Parameters}:
		\begin{itemize}
			\item $M_c, M_r\geq 2$ number of connection and rewiring layers
			\item $q=q_1+\ldots+q_{M_c}$ coordination number \Comment{$q_i\geq 2$}
			\item $r_1, \ldots, r_{M_c+M_r} \in \mathbb{R}^d$ cell displacement vectors
			\item $\ell$ linear cell size \Comment{typically $\ell\approx 8$}
			\item $s$ number of rewiring attempts per cell and layer
			\item $T$ initial SA temperature
		\end{itemize}
		\State \textbf{Complexity}: $\mathcal{O}(sqM_rN)$
		\phase{Initial Connections}
		\For{$a=1$ \textbf{to} $M_c$}\Comment{iterate over connection layers}
		\State subdivide spatial domain into cells $\mathcal{K}_{a,n}$ 
		\State of size $\ell^d$, displaced by $r_a$ w.r.t.~the origin
		\ForAll{cells $\mathcal{K}_{a,n}$}
		\State \textbf{call} \textsc{Connect\_Subgraph}(sites in $\mathcal{K}_{a,n}$; $q_a$)
		\EndFor
		\EndFor
		\phase{Dynamical Rewiring}
		\For{$a=1$ \textbf{to} $M_r$}\Comment{iterate over rewiring layers}
		\State subdivide spatial domain into cells $\mathcal{K}_{a,n}$ 
		\State of size $\ell^d$, displaced by $r_{a+M_c}$ w.r.t.~the origin
		\ForAll{open cells $\tilde{\mathcal{K}}_{a,n}$} 
		\For{$ b=1 $ \textbf{to} $s\cdot q$} \Comment{number of repetitions per cell}
		\State $i$, $j$ $\leftarrow$ two random sites $\in G$
		\State $l$ $\leftarrow$ random neighbor of $i$
		\State $k$ $\leftarrow$ random neighbor of $j$
		\If{\textbf{not} any duplicates in $\{i,j,k,l\}$ \\ \textbf{and} $i$ \textbf{is not} neighbor of $j$ \\
			\textbf{and} $k$ \textbf{is not} neighbor of $l$}
		\State \textbf{call} \textsc{Rewiring\_Attempt}($ {i,j,k,l} $; $ T $)
		\EndIf
		\EndFor
		\EndFor
		\State decrease temperature
		\EndFor
		\State \textbf{repeat} Phase 2 \textbf{until} desired locality is reached
	\end{algorithmic}
	\hrule height \phaserulewidth
	\caption{Pseudocode for the construction of the CC lattice. The subroutines \textsc{Connect\_Subgraph} and \textsc{Rewiring\_Attempt} are described in Fig.~\ref{alg:Helper}.}
	\label{alg:AlgorithmMain}
\end{figure}

In order to eliminate these deficiencies, we impose that all the steps of the lattice construction must be local, restricted to subgraphs delimited by grid cells of a small, fixed size instead of over the whole lattice.
In the first step of the construction (Fig.~\ref{fig:cc_construction}a), the spatial domain of linear size $L$ is subdivided into cells $ \mathcal{K}_{1,n} $ of linear size $\ell$ and index $n$. Typically, we choose $\ell\approx8$. In the next step (Fig.~\ref{fig:cc_construction}b), the sites in each cell are linked together, such that each of them has $q_1<q$ neighbors. Note that building these subgraphs is always possible as long as $q_1$ is even~\footnote{That we can always assign an even number $q_i$ of neighbors to any set of sites can be easily seen by arranging the sites in a circle and connecting every site to its $q_i/2$ left (or right) neighbors.}, a restriction imposed by the \emph{handshaking lemma}~\cite{euler1741}. Once this first \emph{layer} of connections is in place, the lattice consists of $ (L/\ell)^d $ disconnected subgraphs where $d=2$ in Fig.~\ref{fig:cc_construction}. Then, considering a second grid of cells $\mathcal{K}_{2,n}$ that is, for instance, diagonally displaced with respect to $\mathcal{K}_{1,n}$, another set of bonds is added so that each site gets $q_2$ additional bonds (Fig.~\ref{fig:cc_construction}c). This staggering of layers results in a seamless, connected graph where each site has $q_1+q_2$ neighbors. In the next step (Fig.~\ref{fig:cc_construction}d) bond lengths are reduced by a simulated annealing procedure as described in Sec.~\ref{sec:BasicConcept}, but now restricted to the cells $ \mathcal{\tilde{K}}_{1,n} $, $ \mathcal{\tilde{K}}_{2,n} $, \dots where the tilde indicates that the cells are \text{open}, i.e., not only those bonds are considered which lie completely inside the cells, but also those crossing the boundaries. In contrast to the initial connection steps (b, c), the SA step may be repeated not only for diagonally staggered cells (Fig.~\ref{fig:cc_construction}e), but also for cells displaced horizontally or vertically by $\ell/2$ (not shown in the figure). The rewiring is repeated until the desired degree of locality is reached, completing the construction of the CC lattice (Fig.~\ref{fig:cc_construction}f). Note that, in order to avoid any directional bias, we switch the order in which the cells are processed (bottom left to top right in Fig.~\ref{fig:cc_construction}) after each repetition. A detailed pseudocode for the construction procedure can be found in Fig.~\ref{alg:AlgorithmMain} and an animation of the whole lattice construction process is available in the Supplementary Material.

The drawbacks of the earlier algorithm, listed above, are eliminated in the present, improved version by restricting the most expensive construction step, namely the dynamical rewiring, to the small $\mathcal{K}$-cells. This drastically reduces the complexity of the lattice construction from $ \mathcal{O}(N^2) $ to $ \mathcal{O}(N) $, as shown in Fig.~\ref{fig:scaling}. 
The small size of the cells also eliminates the need for an optimized initial bond configuration, further saving computing effort, since a fully random initial linking is then sufficient. Also the issue of pathological motifs (Appendix~\ref{sec:PathologicalMotifs}) is thereby eliminated, since it only arises in the generation of an optimized initial configuration, which is not a necessary step in the improved algorithm.
Finally, the property of micro-scale equivalence is now fulfilled by construction as long as the length $\ell$ is fixed for lattices of different size $L$ in a set of finite-size scaling simulations. That can be shown~\cite{schrauth2018a} by considering the distribution of bond lengths for lattices of different sizes: as the histogram of Fig.~\ref{fig:histogram} shows, the bond length distribution for different values of $L$ coincide perfectly within numerical precision. The concentration of lengths around lower values seen in the figure also gives evidence of the high degree of locality of the lattice.

\begin{figure}
	\hrule height \phaserulewidth
	\begin{algorithmic}
		\Statex
		\Procedure{Connect\_Subgraph}{set of sites $G$; $q$}
		\State \textsc{Shuffle} $G$
		\State $l_G$ $ \leftarrow $ length of G
		\For{$i=1$ \textbf{to} $l_G$}
		\For{$j=1$ \textbf{to} $q/2$}
		\State $k\leftarrow\mod(i+j,l_G) $ \Comment{cyclic connections}
		\State Add bond from site $G(i)$ to site $G(k) $
		\EndFor
		\EndFor
		\State \Return
		\EndProcedure
		\Statex
	\end{algorithmic}
	\begin{algorithmic}
		\Procedure{Rewiring\_Attempt}{sites $i,j,k,l$; $T$}
		\State $d_A$ $\leftarrow$ distance($i,l$) + distance($j,k$) \Comment{respecting p.b.c}
		\State $d_B$ $\leftarrow$ distance($i,j$) + distance($k,l$)
		\State $\Delta H \leftarrow d_B-d_A$ \Comment{cost function, Eq.~\eqref{eq:cost}}
		\State Draw random number $r\in[0,1)$
		\If{$r<\min(1,\mathrm{e}^{-\Delta H / T})$}
		\State Remove bond from $i$ to $l$
		\State Remove bond from $j$ to $k$
		\State Add bond from $i$ to $j$
		\State Add bond from $k$ to $l$
		\EndIf
		\State \Return
		\EndProcedure
	\end{algorithmic}
	\hrule height \phaserulewidth
	\caption{Pseudocode for the subroutines used in Fig.~\ref{alg:AlgorithmMain}.}
	\label{alg:Helper}
\end{figure}

We remark that some care must be taken with the set of construction parameters $q_i$. The handshaking lemma states that any finite graph has an even number of odd-degree nodes. As a consequence, the algorithm cannot converge for cells which end up with an odd number of nodes if $q_i$ is also odd. Clearly, this also places an important limitation on the algorithm: it should only be used for generating lattices with even coordination number $q$. Provided $q_i$ are even, the number of layers and the amount of displacement can be seen as tunable parameters. For constructing a lattice of constant coordination number $q=6$, for instance, one could employ either $q=q_1+q_2=2+4=6$, or three layers with $q=q_1+q_2+q_3=2+2+2=6$, and displacements $ \ell/3 $ and $ 2\ell/3 $. A two-layer setting $q=q_1+q_2=3+3=6$ must be avoided due to the evenness requirement from the handshaking lemma. An overview of some of the possible configurations is given in Tab.~\ref{tab:Displacements}.

\begin{table}[b]
	\centering
	\begin{tabularx}{\columnwidth}{C{3mm}C{4mm}C{18mm}X} 
		\hline
		\hline
		$d$&$q$ & $q_i$ & displacements \Vghost\\
		\hline
		2&4 & 2+2     & $(0,0)$, $(\ell/2,\ell/2)$ \Vghost\\
		2&6 & 2+2+2   & $(0,0)$, $(\ell/3,\ell/3)$, $(2\ell/3,2\ell/3)$ \\
		2&6 & 4+2     & $(0,0)$, $(\ell/2,\ell/2)$ \\
		2&8 & 4+4     & $(0,0)$, $(\ell/2,\ell/2)$ \\
		2&8 & 2+2+2+2 & $(n\ell/4,n\ell/4)$, $n=0,1,2,3$ \\
		2&8 & 2+2+2+2 & $(0,0)$, $(\ell/2,0)$, $(0,\ell/2)$, $(\ell/2,\ell/2)$ \\
		\hline
		3&4 & 2+2     & $(0,0,0)$, $(\ell/2,\ell/2,\ell/2)$ \Vghost\\
		3&6 & 2+2+2   & $(n\ell/3,n\ell/3,n\ell/3)$, $n=0,1,2$ \\
		3&8 & 2+2+2+2 & $(n\ell/4,n\ell/4,n\ell/4)$, $n=0,1,2,3$ \\
		3&8 & 2+2+2+2 & $(0,0,0)$, $(\ell/2,0,0)$, $(0,\ell/2,0)$, $(0,0,\ell/2)$ \\
		\hline \hline
	\end{tabularx} 
	\caption{Examples of possible coordination number decompositions and corresponding cell displacements of the initial connection layers for two and three dimensions.}
	\label{tab:Displacements}
\end{table}
%

\section{Higher Dimensions}
\label{sec:HigherDimensions}

Whereas the $\mathcal{O}(N^2)$ scaling of the original algorithm is a significant limitation already in 2D, for higher dimensions it makes the construction of lattices of reasonable size prohibitively expensive. In this context, it is important to notice that the current algorithm is not only a substantial improvement over the original one, but also compares favorably with algorithms for usual proximity graphs, such as the Delaunay triangulation and its subgraphs, as well as nearest-neighbor graphs. For the latter, typical sequential algorithms are known to scale as $ \mathcal{O}(N\ln N) $, through the use of spatial tree decomposition methods~\cite{su1997,supowit1983,vaidya1989}. For the DT on uniformly distributed points, a $ \mathcal{O}(N\ln\ln N) $ scaling is possible~\cite{dwyer1987} using sophisticated divide-and-conquer techniques. In dimensions larger than two, some of those algorithms are not trivially generalized and the scaling is not maintained, such as for the RNG, where one falls back to algebraic complexity in 3D~\cite{agarwal1992}. In contrast, the CC algorithm is straightforwardly generalized to higher dimensions, as already described in Fig.~\ref{alg:AlgorithmMain}, while maintaining the $\mathcal{O}(N)$ scaling behavior, as shown in Fig.~\ref{fig:scaling}. Essentially, the construction remains the same, with only some parameters such as the displacement vectors having to be adjusted (see Tab.~\ref{tab:Displacements} for examples).

The hypercubic cells of the general case admit many more layer configurations and displacement vectors than the square cells of the 2D setting (see again Tab.~\ref{tab:Displacements}). One must only ensure that mixing in all directions takes place, such as can always be achieved by a single fully diagonal displacement, i.e.,
\begin{align*}
r_1 = (0,0,\ldots,0), \quad r_2 = (\ell/2,\ell/2,\ldots,\ell/2).
\end{align*}
Hence, as in 2D, the smallest possible coordination number is $q=4$ for any dimension.

Furthermore, the CC algorithm can not only be generalized to higher dimensions, but in principle also to spaces equipped with metrics other than Euclidean, with the single necessary change being in the distance function in the algorithm's rewiring subroutine (Fig.~\ref{alg:Helper}). A generalization to curved manifolds should be possible as long as a proper spatial grid can be defined, such as for the hyperbolic plane $ \mathbb{H}^2 $, for which a number of regular tessellations can be constructed~\cite{dunham1986}. However, due to the inherent length scale of this space (the curvature radius), the cell size is determined by geometric constraints and can not be freely chosen. Also, setting up periodic boundary conditions is non-trivial in hyperbolic spaces, although possible~\cite{sausset2007}.

\section{Application}
\label{sec:Application}

Topologically disordered structures are instrumental in understanding the role of certain perturbations in the context of statistical physics, which motivated the construction of the CC lattice in the first place. Therefore, as a possible application, we perform large-scale Monte-Carlo simulations of the ferromagnetic Ising model on a 3D constant coordination lattice with coordination number $q=4$ (CC4), using the \textsc{Marqov} code framework~\cite{marqov}.

The Ising model is defined by the Hamiltonian
\begin{align}
	\mathcal{H} = -\sum \limits_{ \langle i,j \rangle}J_{ij} s_i s_j +\sum \limits_{i}h_is_i , \quad\quad s_i = \pm 1
	\label{eq:Hamiltonian}
\end{align}
where $s_i$ are discrete spins on the lattice. $J_{ij}$ denotes the coupling between nearest neighbors $\langle i,j \rangle$ and $h_i$ is the external field at site $s_i$. For equilibrium lattice models, quenched disorder can be introduced in a variety of ways. For fixed ferromagnetic coupling $J_{ij} = J >0 $ and randomly distributed external field, the system is called Random Field Ising model (RFIM) and has been investigated thoroughly over the last decades~\cite{nattermann1998}. In contrast, for vanishing external field but randomly distributed (anti)ferromagnetic bonds the system shows the behavior of a spin glass~\cite[and references therein]{binder1986}.

In the present study, quenched disorder is introduced as topological randomness encoded in the implicit connectivity of the CC lattice. We therefore fix all couplings to $J_{ij}=1$ at vanishing external field, $h=0$.

In order to investigate the Ising model in the vicinity of the critical point, we employ state-of-the-art importance-sampling Monte Carlo methods, using cluster, as well as local update algorithms. In particular, we use the algorithm proposed by Wolff \cite{wolff1989}, which significantly reduces the critical slowing down near the critical point and is furthermore straightforwardly applicable to disordered lattices. Although the cluster updates preserve ergodicity, we add local Metropolis updates~\cite{metropolis1953} in order to make sure that the short-wavelength modes are properly thermalized. After an initial thermalization period, magnetization $m$ and energy $e$ per site are measured after every few updates.  

In the study of disordered systems, it is necessary to average physical observables over many different, independent disorder realizations of the system, also called \emph{replicas}. The quenched averages over $N_r$ replicas are performed at the level of (extensive) observables, rather than at the level of the partition function~\cite{binder1986}. Denoting quenched averages as
\begin{align}
	\label{eq:disorder_average}
	[\mathcal{O}]_\text{avg} \equiv \frac{1}{N_r} \sum \limits_{i=1}^{N_r} \mathcal{O}_i
\end{align}
and thermal averages as $\langle ... \rangle $,
we use the following definitions of magnetization, energy and susceptibility:
\begin{subequations}
	\begin{align}
		M &= [ \la |m| \ra ]_\text{avg}, \\
		E &= [ \la e \ra ]_\text{avg}, \\
		\label{eq:obs_chi}
		\chi &= N \beta [ \la m^2 \ra - \la |m| \ra^2  ]_\text{avg}.
	\end{align}
\end{subequations}
Furthermore, the two-point finite-size correlation function is given by
\begin{align}
	\xi =\dfrac{1}{2\sin(k_\text{min}/2)} \sqrt{\dfrac{[\large\langle\large|\mathcal{F}^2(\mathbf{0})\large|\large\rangle]_\text{avg}}{[\large\langle|\mathcal{F}^2(\mathbf{k}_\text{min})\large|\large\rangle]_\text{avg}}-1},
\end{align}
with the Fourier transform of the magnetization defined by
\begin{align}
	\mathcal{F}(\mathbf{\mathbf{k}}) = \sum_{j} s_j\exp(i\mathbf{kx}_j),
\end{align}
where $\mathbf{x}_j$ denotes the spatial coordinate of site $j$ and $  \mathbf{k}_\text{min}=(2\pi/L,0,0) $ represents the smallest non-zero wave vector in the finite system. The quantity $ \mathcal{F}^2(\mathbf{k}_\text{min}) $ is also measured during the Monte Carlo run. Finally, the fourth-order magnetic cumulant, or Binder ratio, is given by
\begin{align}
	U_4 &= \Bigg[ 1 - \frac{ \la m^4 \ra }{ 3 \la m^2 \ra^2}  \Bigg]_\text{avg}.
\end{align}

Essential for the analysis of the Ising model is a precise knowledge of the location of the critical point, which depends on the details of the lattice structure and is therefore not known in advance. It is, however, known that the curves of certain RG-invariant quantities, such as the Binder ratio $U_4$ or the correlation length $\xi/L$ intersect close to the critical point for different lattice sizes $L$. More specifically, the intersection points $ T^* $ for pairs of $(L,2L)$ converge to the critical temperature according to
\begin{align}
T^*(L,2L) = T_c+aL^{-1/\nu},
\label{eq:CriticalTemperature}
\end{align}
which allows to obtain a very precise estimate of the critical temperature~\cite{hasenbusch2010}. In the scope of this paper, however, we simply use the crossings without the infinite volume extrapolation, which provides a sufficiently good estimate for the next steps of the analysis. From the $U_4$ curves we get the following estimate
\begin{align}
T_c = 2.4818(2),
\label{eq:T_c}
\end{align}
where the error is evaluated, quite conservatively, from the width of the intersection region. The estimate from the crossing points of $\xi/L$ turns out to be considerably less precise, though fully compatible. For the fixed point values of the phenomenological quantities, we obtain
\begin{align}
(\xi/L)^* &= 0.623(10),\\
U_4^* &= 0.468(4),
\end{align}
again without using infinite volume extrapolations. These quantities are considered universal, at least in a limited sense, in that they depend weakly on certain geometrical characteristics of the system \cite{selke2005,selke2009,malakis2014}. Taking as reference the most precise estimates available for the 3D Ising model on a cubic lattice, $ (\xi/L)^* = 0.6431(1) $~\cite{hasenbusch2010} and $ U_4^*=0.46548(5) $~\cite{ferrenberg2018}, we see that our estimates present only small deviations, giving a first indication that the Ising model on a 3D CC4 lattice stays in the universality class of the clean model.

In the next step of the analysis we simulate the Ising model on lattices of size $L=16,24,32,48,64,96,$ and $128$ for several temperatures in the vicinity of the critical point. A measurement is taken after every Elementary Monte Carlo Step (EMCS), which consists of one Metropolis sweep and $L$ cluster updates, keeping the fraction of flipped sites approximately independent of the lattice size. Each disorder realization is initially prepared in a cold configuration and 500 EMCS are used for proper equilibration. Then another 1000 EMCS are performed, with a measurement being taken after each one of them. For the smaller lattices we use up to $10^4$ disorder replicas for the averages of Eq.~\eqref{eq:disorder_average}, for the two largest lattice, $L=96$ and $L=128$, we use at least 1500 replicas for every temperature. The simulations took about $10^4$ CPU days on an Intel Xeon E5-2697 v3 processor, where the time for constructing the lattices was below 1\% of the total time.

\begin{figure}
	\centering
	\includegraphics[width=\columnwidth]{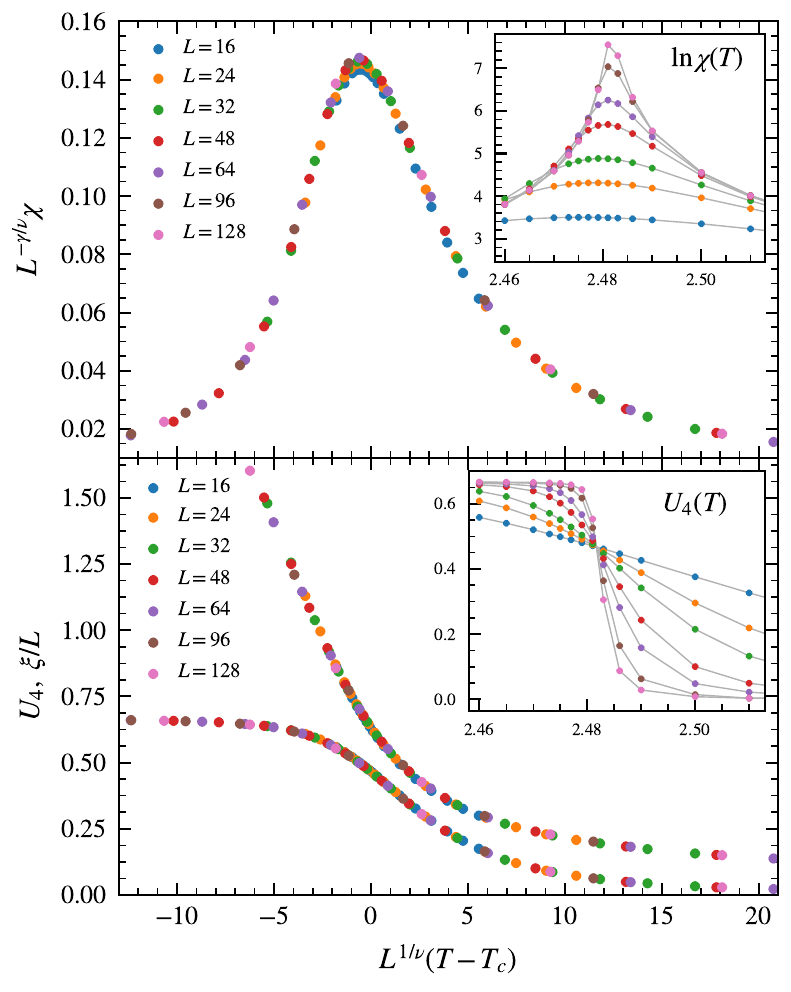}
	\caption{[Color online] Upper panel: Finite-size data collapse of the magnetic susceptibility according to Eq.~\ref{eq:scaling_relations}. Lower panel: Data collapse for the fourth-order Binder cumulent (lower curve) and two-point finite-size correlation function (upper curve). Insets show the non-rescaled observables. The gray lines are guides to the eye.}
	\label{fig:ising}		
\end{figure}

Finite-size scaling theory predicts that the susceptibility, the fourth-order magnetic cumulant and the correlation length scale according to 
\begin{subequations}
	\begin{align}
	\chi &= L^{\gamma/\nu} f_\chi(x)\,( 1 + \ldots ), \\
	U_4 &= f_{U_4}(x)\,( 1 + \ldots ), \\
	\xi/L &= f_\xi(x)\,( 1 + \ldots ),
	\label{eq:scaling_relations}
\end{align}
\end{subequations}
where $\gamma$ and $\nu$ are critical exponents and $f_\chi$, $f_{U_4}$, and $f_\xi$ denote universal scaling functions, with the argument $x$ given by
\begin{align}
	x = ( T - T_c)L^{1/\nu}.
\end{align}
These equations describe the scaling behavior to first order. Corrections of higher order are expected to become irrelevant for large system sizes. We compute the scaling collapse plots, assuming the best known values for the clean model critical exponents~\cite{ferrenberg2018} and the value of $T_c$ of Eq.~\eqref{eq:T_c}. In the upper panel of Fig.~\ref{fig:ising} we plot the scaling collapse for the susceptibility and, in its lower panel we plot both phenomenological couplings against the scaling variable. The nearly flawless collapse displayed by the curves, for lattice sizes $L\geq32$ in the upper panel and for $ L\geq 16 $ in the lower panel, provides compelling evidence that the Ising model on a three-dimensional CC4 lattice belongs to the universality class of the clean 3D Ising model.

\section{Conclusion}
\label{sec:Conclusion}

We present an algorithm for the construction of a topologically disordered lattice with nodes of constant coordination number. Boasting a computational complexity that scales linearly with the number of points, it is significantly faster than other proximity graph constructions. The algorithm performs only local operations, dividing the spatial domain into cells of small linear size compared to the lattice dimensions. This guarantees bond lengths are bounded and allows a straightforward generalization to any number of spatial dimensions and, in principle, to different metrics and topologies. By efficiently constructing disordered graphs of fixed coordinated number, the CC lattice could find application in the modeling of amorphous materials such as low temperature amorphous silicon in two and three dimensions~\cite{keblinski2002, buchner2017}, especially given that the most well-known lattice with constant coordination number, the Voronoi construction, is not considered a satisfactory model \cite{gaspard1985}.
As a prototypical application we perform numerical Monte Carlo simulations of the 3D equilibrium Ising model on a realization of the lattice with four neighbors and find the character of the second-order phase transition to be unchanged with respect to the clean 3D Ising model. Hence, quenched disorder is revealed to be a non-relevant perturbation in this case, providing another puzzle piece on the road towards a general disorder relevance criterion~\cite{schrauth2018b}.

\begin{acknowledgments}
	We thank H.~Hinrichsen and F.~Goth for helpful discussions. M.S.~thanks the Studienstiftung des deutschen Volkes for financial support. This work is part of the DFG research project Hi~744/9-1. The authors gratefully acknowledge the Gauss Centre for Supercomputing e.V. (www.gauss-centre.eu) for funding this project by providing computing time on the GCS Supercomputer SuperMUC at Leibniz Supercomputing Centre (www.lrz.de).
\end{acknowledgments}

\setcounter{table}{0}
\setcounter{figure}{0}   
\setcounter{equation}{0}   
\renewcommand{\thetable}{\thesection.\arabic{table}}
\renewcommand{\thefigure}{\thesection.\arabic{figure}}
\renewcommand{\theequation}{\thesection.\arabic{equation}}
\appendix

\section{Pathological Motifs}
\label{sec:PathologicalMotifs}

\begin{figure}[h]
	\centering      
	\includegraphics[width=0.95\linewidth]{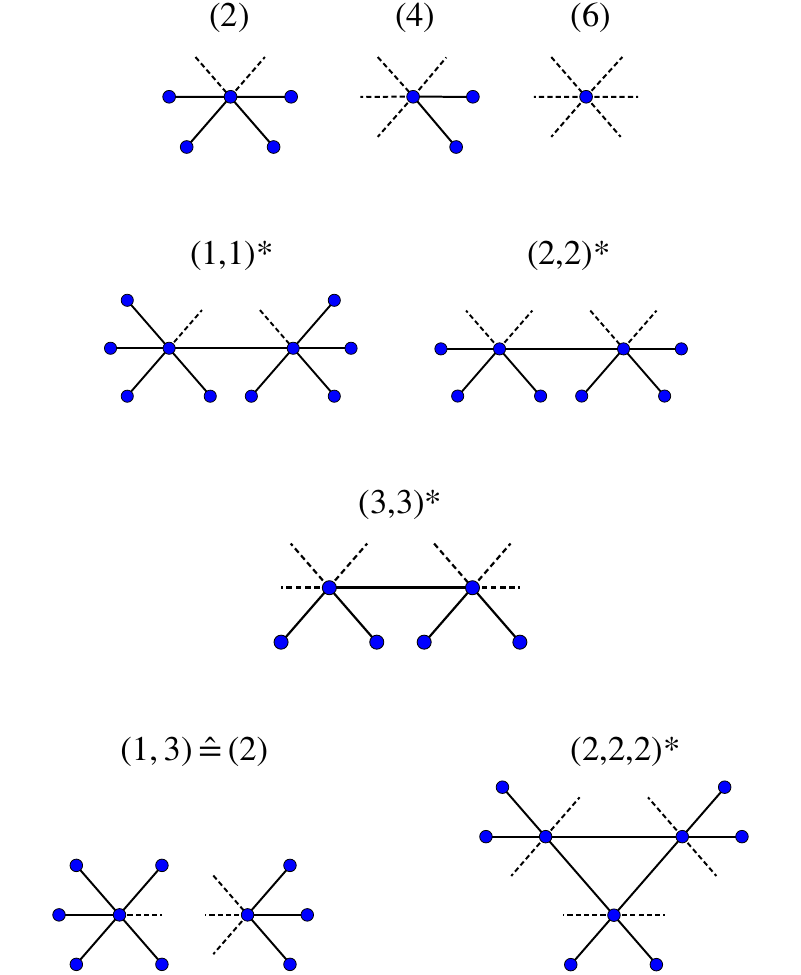}
	\caption{Pathological motifs which may appear during the initial construction step of the originally proposed CC lattice algorithm. Dashed lines symbolize open connectors (which form a bond when two are closed/connected) at the end of the initial construction step, whereas the peripheral blue dots represent sites that are already fully connected.}
	\label{fig:pathological-motifs}         
\end{figure}

Fig.~\ref{fig:pathological-motifs} shows the most frequently encountered ``pathological'' or ``degenerate'' motifs which may occur during the initial construction step of the originally proposed CC algorithm and cause it to fail.  In the case labeled by $ (2) $, for instance, every site except for one is already fully connected. This site, however, still has two dangling connectors, which clearly would lead to a self-connection, which is illegal. Another example is the case $ (1,1)^* $, where eventually two sites remain with one dangling bond each. As they, however, are already connected, this would result in a double-connection, which is also illegal. In the improved CC algorithm, presented in this paper, these issue do not arise in the first place, as fully random initial connections are sufficient, as pointed out in Sec.~\ref{sec:AlgorithmicDetails}.

\begin{thebibliography}{56}%
\makeatletter
\providecommand \@ifxundefined [1]{%
 \@ifx{#1\undefined}
}%
\providecommand \@ifnum [1]{%
 \ifnum #1\expandafter \@firstoftwo
 \else \expandafter \@secondoftwo
 \fi
}%
\providecommand \@ifx [1]{%
 \ifx #1\expandafter \@firstoftwo
 \else \expandafter \@secondoftwo
 \fi
}%
\providecommand \natexlab [1]{#1}%
\providecommand \enquote  [1]{``#1''}%
\providecommand \bibnamefont  [1]{#1}%
\providecommand \bibfnamefont [1]{#1}%
\providecommand \citenamefont [1]{#1}%
\providecommand \href@noop [0]{\@secondoftwo}%
\providecommand \href [0]{\begingroup \@sanitize@url \@href}%
\providecommand \@href[1]{\@@startlink{#1}\@@href}%
\providecommand \@@href[1]{\endgroup#1\@@endlink}%
\providecommand \@sanitize@url [0]{\catcode `\\12\catcode `\$12\catcode
  `\&12\catcode `\#12\catcode `\^12\catcode `\_12\catcode `\%12\relax}%
\providecommand \@@startlink[1]{}%
\providecommand \@@endlink[0]{}%
\providecommand \url  [0]{\begingroup\@sanitize@url \@url }%
\providecommand \@url [1]{\endgroup\@href {#1}{\urlprefix }}%
\providecommand \urlprefix  [0]{URL }%
\providecommand \Eprint [0]{\href }%
\providecommand \doibase [0]{http://dx.doi.org/}%
\providecommand \selectlanguage [0]{\@gobble}%
\providecommand \bibinfo  [0]{\@secondoftwo}%
\providecommand \bibfield  [0]{\@secondoftwo}%
\providecommand \translation [1]{[#1]}%
\providecommand \BibitemOpen [0]{}%
\providecommand \bibitemStop [0]{}%
\providecommand \bibitemNoStop [0]{.\EOS\space}%
\providecommand \EOS [0]{\spacefactor3000\relax}%
\providecommand \BibitemShut  [1]{\csname bibitem#1\endcsname}%
\let\auto@bib@innerbib\@empty
\bibitem [{\citenamefont {Watts}\ and\ \citenamefont
  {Strogatz}(1998)}]{watts1998}%
  \BibitemOpen
  \bibfield  {author} {\bibinfo {author} {\bibfnamefont {D.~J.}\ \bibnamefont
  {Watts}}\ and\ \bibinfo {author} {\bibfnamefont {S.~H.}\ \bibnamefont
  {Strogatz}},\ }\href@noop {} {\bibfield  {journal} {\bibinfo  {journal}
  {Nature}\ }\textbf {\bibinfo {volume} {393}},\ \bibinfo {pages} {440}
  (\bibinfo {year} {1998})}\BibitemShut {NoStop}%
\bibitem [{\citenamefont {Cohen}\ and\ \citenamefont
  {Havlin}(2003)}]{cohen2003}%
  \BibitemOpen
  \bibfield  {author} {\bibinfo {author} {\bibfnamefont {R.}~\bibnamefont
  {Cohen}}\ and\ \bibinfo {author} {\bibfnamefont {S.}~\bibnamefont {Havlin}},\
  }\href@noop {} {\bibfield  {journal} {\bibinfo  {journal} {Phys. Rev. Lett.}\
  }\textbf {\bibinfo {volume} {90}},\ \bibinfo {pages} {058701} (\bibinfo
  {year} {2003})}\BibitemShut {NoStop}%
\bibitem [{\citenamefont {Toth}\ \emph {et~al.}(2017)\citenamefont {Toth},
  \citenamefont {O'Rourke},\ and\ \citenamefont {Goodman}}]{toth2017}%
  \BibitemOpen
  \bibfield  {author} {\bibinfo {author} {\bibfnamefont {C.~D.}\ \bibnamefont
  {Toth}}, \bibinfo {author} {\bibfnamefont {J.}~\bibnamefont {O'Rourke}}, \
  and\ \bibinfo {author} {\bibfnamefont {J.~E.}\ \bibnamefont {Goodman}},\
  }\href@noop {} {\emph {\bibinfo {title} {Handbook of {D}iscrete and
  {C}omputational {G}eometry}}}\ (\bibinfo  {publisher} {Chapman and
  Hall/CRC},\ \bibinfo {year} {2017})\BibitemShut {NoStop}%
\bibitem [{\citenamefont {Okabe}\ \emph {et~al.}(2000)\citenamefont {Okabe},
  \citenamefont {Boots}, \citenamefont {Sugiharaa},\ and\ \citenamefont
  {Chiu}}]{okabe2000}%
  \BibitemOpen
  \bibfield  {author} {\bibinfo {author} {\bibfnamefont {A.}~\bibnamefont
  {Okabe}}, \bibinfo {author} {\bibfnamefont {B.}~\bibnamefont {Boots}},
  \bibinfo {author} {\bibfnamefont {K.}~\bibnamefont {Sugiharaa}}, \ and\
  \bibinfo {author} {\bibfnamefont {S.~N.}\ \bibnamefont {Chiu}},\ }\href@noop
  {} {\emph {\bibinfo {title} {{S}patial {T}essellations: {C}oncepts and
  {A}pplications of {V}oronoi {D}iagrams}}},\ \bibinfo {edition} {2nd}\ ed.\
  (\bibinfo  {publisher} {Wiley},\ \bibinfo {address} {Chichester},\ \bibinfo
  {year} {2000})\BibitemShut {NoStop}%
\bibitem [{\citenamefont {Toussaint}(1980)}]{toussaint1980}%
  \BibitemOpen
  \bibfield  {author} {\bibinfo {author} {\bibfnamefont {G.~T.}\ \bibnamefont
  {Toussaint}},\ }\href@noop {} {\bibfield  {journal} {\bibinfo  {journal}
  {Pattern Recognit.}\ }\textbf {\bibinfo {volume} {12}},\ \bibinfo {pages}
  {261} (\bibinfo {year} {1980})}\BibitemShut {NoStop}%
\bibitem [{\citenamefont {Gabriel}\ and\ \citenamefont
  {Sokal}(1969)}]{gabriel1969}%
  \BibitemOpen
  \bibfield  {author} {\bibinfo {author} {\bibfnamefont {K.~R.}\ \bibnamefont
  {Gabriel}}\ and\ \bibinfo {author} {\bibfnamefont {R.~R.}\ \bibnamefont
  {Sokal}},\ }\href@noop {} {\bibfield  {journal} {\bibinfo  {journal}
  {Systematic Zoology}\ }\textbf {\bibinfo {volume} {18}},\ \bibinfo {pages}
  {259} (\bibinfo {year} {1969})}\BibitemShut {NoStop}%
\bibitem [{\citenamefont {Jaromczyk}\ and\ \citenamefont
  {Toussaint}(1992)}]{jaromczyk1992}%
  \BibitemOpen
  \bibfield  {author} {\bibinfo {author} {\bibfnamefont {J.~W.}\ \bibnamefont
  {Jaromczyk}}\ and\ \bibinfo {author} {\bibfnamefont {G.~T.}\ \bibnamefont
  {Toussaint}},\ }\href@noop {} {\bibfield  {journal} {\bibinfo  {journal}
  {Proceedings of the IEEE}\ }\textbf {\bibinfo {volume} {80}},\ \bibinfo
  {pages} {1502} (\bibinfo {year} {1992})}\BibitemShut {NoStop}%
\bibitem [{\citenamefont {Barth{\'e}lemy}(2011)}]{barthelemy2011}%
  \BibitemOpen
  \bibfield  {author} {\bibinfo {author} {\bibfnamefont {M.}~\bibnamefont
  {Barth{\'e}lemy}},\ }\href@noop {} {\bibfield  {journal} {\bibinfo  {journal}
  {Phys.~Rep.}\ }\textbf {\bibinfo {volume} {499}},\ \bibinfo {pages} {1}
  (\bibinfo {year} {2011})}\BibitemShut {NoStop}%
\bibitem [{\citenamefont {Aurenhammer}(1991)}]{aurenhammer1991}%
  \BibitemOpen
  \bibfield  {author} {\bibinfo {author} {\bibfnamefont {F.}~\bibnamefont
  {Aurenhammer}},\ }\href@noop {} {\bibfield  {journal} {\bibinfo  {journal}
  {ACM Computing Surveys (CSUR)}\ }\textbf {\bibinfo {volume} {23}},\ \bibinfo
  {pages} {345} (\bibinfo {year} {1991})}\BibitemShut {NoStop}%
\bibitem [{\citenamefont {Weatherill}(1992)}]{weatherill1992}%
  \BibitemOpen
  \bibfield  {author} {\bibinfo {author} {\bibfnamefont {N.}~\bibnamefont
  {Weatherill}},\ }\href {\doibase
  https://doi.org/10.1016/0898-1221(92)90045-J} {\bibfield  {journal} {\bibinfo
   {journal} {Comput. Math. Appl.}\ }\textbf {\bibinfo {volume} {24}},\
  \bibinfo {pages} {129 } (\bibinfo {year} {1992})}\BibitemShut {NoStop}%
\bibitem [{\citenamefont {{Bebis}}\ \emph {et~al.}(1999)\citenamefont
  {{Bebis}}, \citenamefont {{Deaconu}},\ and\ \citenamefont
  {{Georgiopoulos}}}]{bebis1999}%
  \BibitemOpen
  \bibfield  {author} {\bibinfo {author} {\bibfnamefont {G.}~\bibnamefont
  {{Bebis}}}, \bibinfo {author} {\bibfnamefont {T.}~\bibnamefont {{Deaconu}}},
  \ and\ \bibinfo {author} {\bibfnamefont {M.}~\bibnamefont
  {{Georgiopoulos}}},\ }in\ \href {\doibase 10.1109/ICIIS.1999.810315} {\emph
  {\bibinfo {booktitle} {Proceedings 1999 International Conference on
  Information Intelligence and Systems}}}\ (\bibinfo {year} {1999})\ pp.\
  \bibinfo {pages} {452--459}\BibitemShut {NoStop}%
\bibitem [{\citenamefont {Gopi}\ \emph {et~al.}(2000)\citenamefont {Gopi},
  \citenamefont {Krishnan},\ and\ \citenamefont {Silva}}]{gopi2000}%
  \BibitemOpen
  \bibfield  {author} {\bibinfo {author} {\bibfnamefont {M.}~\bibnamefont
  {Gopi}}, \bibinfo {author} {\bibfnamefont {S.}~\bibnamefont {Krishnan}}, \
  and\ \bibinfo {author} {\bibfnamefont {C.~T.}\ \bibnamefont {Silva}},\ }in\
  \href@noop {} {\emph {\bibinfo {booktitle} {Computer Graphics Forum}}},\
  Vol.~\bibinfo {volume} {19}\ (\bibinfo {organization} {Wiley Online
  Library},\ \bibinfo {year} {2000})\ pp.\ \bibinfo {pages}
  {467--478}\BibitemShut {NoStop}%
\bibitem [{\citenamefont {Ramella}\ \emph {et~al.}(2001)\citenamefont
  {Ramella}, \citenamefont {Boschin}, \citenamefont {Fadda},\ and\
  \citenamefont {Nonino}}]{ramella2001}%
  \BibitemOpen
  \bibfield  {author} {\bibinfo {author} {\bibfnamefont {M.}~\bibnamefont
  {Ramella}}, \bibinfo {author} {\bibfnamefont {W.}~\bibnamefont {Boschin}},
  \bibinfo {author} {\bibfnamefont {D.}~\bibnamefont {Fadda}}, \ and\ \bibinfo
  {author} {\bibfnamefont {M.}~\bibnamefont {Nonino}},\ }\href@noop {}
  {\bibfield  {journal} {\bibinfo  {journal} {Astronomy \& Astrophysics}\
  }\textbf {\bibinfo {volume} {368}},\ \bibinfo {pages} {776} (\bibinfo {year}
  {2001})}\BibitemShut {NoStop}%
\bibitem [{\citenamefont {Wu}\ \emph {et~al.}(2007)\citenamefont {Wu},
  \citenamefont {Lee},\ and\ \citenamefont {Chung}}]{wu2007}%
  \BibitemOpen
  \bibfield  {author} {\bibinfo {author} {\bibfnamefont {C.-H.}\ \bibnamefont
  {Wu}}, \bibinfo {author} {\bibfnamefont {K.-C.}\ \bibnamefont {Lee}}, \ and\
  \bibinfo {author} {\bibfnamefont {Y.-C.}\ \bibnamefont {Chung}},\ }\href
  {\doibase https://doi.org/10.1016/j.comcom.2007.05.017} {\bibfield  {journal}
  {\bibinfo  {journal} {Comput. Commun.}\ }\textbf {\bibinfo {volume} {30}},\
  \bibinfo {pages} {2744 } (\bibinfo {year} {2007})}\BibitemShut {NoStop}%
\bibitem [{\citenamefont {Chandler}(1987)}]{chandler1987}%
  \BibitemOpen
  \bibfield  {author} {\bibinfo {author} {\bibfnamefont {D.}~\bibnamefont
  {Chandler}},\ }\href@noop {} {\emph {\bibinfo {title} {{I}ntroduction to
  {M}odern {S}tatistical {M}echanics}}}\ (\bibinfo  {publisher} {Oxford
  University Press Inc},\ \bibinfo {year} {1987})\BibitemShut {NoStop}%
\bibitem [{\citenamefont {T{\"a}uber}(2014)}]{tauber2014}%
  \BibitemOpen
  \bibfield  {author} {\bibinfo {author} {\bibfnamefont {U.~C.}\ \bibnamefont
  {T{\"a}uber}},\ }\href@noop {} {\emph {\bibinfo {title} {{C}ritical
  {D}ynamics: {A} {F}ield {T}heory {A}pproach to {E}quilibrium and
  {N}on-{E}quilibrium {S}caling {B}ehavior}}}\ (\bibinfo  {publisher}
  {Cambridge University Press},\ \bibinfo {year} {2014})\BibitemShut {NoStop}%
\bibitem [{\citenamefont {Martins}\ and\ \citenamefont
  {Plascak}(2007)}]{martins2007}%
  \BibitemOpen
  \bibfield  {author} {\bibinfo {author} {\bibfnamefont {P.~H.~L.}\
  \bibnamefont {Martins}}\ and\ \bibinfo {author} {\bibfnamefont {J.~A.}\
  \bibnamefont {Plascak}},\ }\href {\doibase 10.1103/PhysRevE.76.012102}
  {\bibfield  {journal} {\bibinfo  {journal} {Phys. Rev. E}\ }\textbf {\bibinfo
  {volume} {76}},\ \bibinfo {pages} {012102} (\bibinfo {year}
  {2007})}\BibitemShut {NoStop}%
\bibitem [{\citenamefont {Pelissetto}\ and\ \citenamefont
  {Vicari}(2002)}]{pelissetto2002}%
  \BibitemOpen
  \bibfield  {author} {\bibinfo {author} {\bibfnamefont {A.}~\bibnamefont
  {Pelissetto}}\ and\ \bibinfo {author} {\bibfnamefont {E.}~\bibnamefont
  {Vicari}},\ }\href@noop {} {\bibfield  {journal} {\bibinfo  {journal}
  {Physics Reports}\ }\textbf {\bibinfo {volume} {368}},\ \bibinfo {pages}
  {549} (\bibinfo {year} {2002})}\BibitemShut {NoStop}%
\bibitem [{\citenamefont {Ising}(1925)}]{ising1925}%
  \BibitemOpen
  \bibfield  {author} {\bibinfo {author} {\bibfnamefont {E.}~\bibnamefont
  {Ising}},\ }\href@noop {} {\bibfield  {journal} {\bibinfo  {journal}
  {Zeitschrift f{ü}r Physik A}\ }\textbf {\bibinfo {volume} {31}},\ \bibinfo
  {pages} {253} (\bibinfo {year} {1925})}\BibitemShut {NoStop}%
\bibitem [{\citenamefont {Hinrichsen}(2000)}]{hinrichsen2000}%
  \BibitemOpen
  \bibfield  {author} {\bibinfo {author} {\bibfnamefont {H.}~\bibnamefont
  {Hinrichsen}},\ }\href {\doibase 10.1080/00018730050198152} {\bibfield
  {journal} {\bibinfo  {journal} {Advances in Physics}\ }\textbf {\bibinfo
  {volume} {49}},\ \bibinfo {pages} {815} (\bibinfo {year} {2000})}\BibitemShut
  {NoStop}%
\bibitem [{\citenamefont {Janke}\ \emph {et~al.}(1993)\citenamefont {Janke},
  \citenamefont {Katoot},\ and\ \citenamefont {Villanova}}]{janke1993}%
  \BibitemOpen
  \bibfield  {author} {\bibinfo {author} {\bibfnamefont {W.}~\bibnamefont
  {Janke}}, \bibinfo {author} {\bibfnamefont {M.}~\bibnamefont {Katoot}}, \
  and\ \bibinfo {author} {\bibfnamefont {R.}~\bibnamefont {Villanova}},\ }\href
  {\doibase http://dx.doi.org/10.1016/0370-2693(93)91633-X} {\bibfield
  {journal} {\bibinfo  {journal} {Phys. Lett. B}\ }\textbf {\bibinfo {volume}
  {315}},\ \bibinfo {pages} {412 } (\bibinfo {year} {1993})}\BibitemShut
  {NoStop}%
\bibitem [{\citenamefont {Janke}\ \emph {et~al.}(1994)\citenamefont {Janke},
  \citenamefont {Katoot},\ and\ \citenamefont {Villanova}}]{janke1994}%
  \BibitemOpen
  \bibfield  {author} {\bibinfo {author} {\bibfnamefont {W.}~\bibnamefont
  {Janke}}, \bibinfo {author} {\bibfnamefont {M.}~\bibnamefont {Katoot}}, \
  and\ \bibinfo {author} {\bibfnamefont {R.}~\bibnamefont {Villanova}},\ }\href
  {\doibase 10.1103/PhysRevB.49.9644} {\bibfield  {journal} {\bibinfo
  {journal} {Phys. Rev. B}\ }\textbf {\bibinfo {volume} {49}},\ \bibinfo
  {pages} {9644} (\bibinfo {year} {1994})}\BibitemShut {NoStop}%
\bibitem [{\citenamefont {Lima}\ \emph {et~al.}(2000)\citenamefont {Lima},
  \citenamefont {Moreira}, \citenamefont {Andrade},\ and\ \citenamefont
  {Costa}}]{lima2000}%
  \BibitemOpen
  \bibfield  {author} {\bibinfo {author} {\bibfnamefont {F.}~\bibnamefont
  {Lima}}, \bibinfo {author} {\bibfnamefont {J.}~\bibnamefont {Moreira}},
  \bibinfo {author} {\bibfnamefont {J.}~\bibnamefont {Andrade}}, \ and\
  \bibinfo {author} {\bibfnamefont {U.}~\bibnamefont {Costa}},\ }\href
  {\doibase http://dx.doi.org/10.1016/S0378-4371(00)00134-5} {\bibfield
  {journal} {\bibinfo  {journal} {Physica A}\ }\textbf {\bibinfo {volume}
  {283}},\ \bibinfo {pages} {100 } (\bibinfo {year} {2000})}\BibitemShut
  {NoStop}%
\bibitem [{\citenamefont {Janke}\ and\ \citenamefont
  {Villanova}(2002)}]{janke2002}%
  \BibitemOpen
  \bibfield  {author} {\bibinfo {author} {\bibfnamefont {W.}~\bibnamefont
  {Janke}}\ and\ \bibinfo {author} {\bibfnamefont {R.}~\bibnamefont
  {Villanova}},\ }\href {\doibase 10.1103/PhysRevB.66.134208} {\bibfield
  {journal} {\bibinfo  {journal} {Phys. Rev. B}\ }\textbf {\bibinfo {volume}
  {66}},\ \bibinfo {pages} {134208} (\bibinfo {year} {2002})}\BibitemShut
  {NoStop}%
\bibitem [{\citenamefont {de~Oliveira}\ \emph {et~al.}(2008)\citenamefont
  {de~Oliveira}, \citenamefont {Alves}, \citenamefont {Ferreira},\ and\
  \citenamefont {Dickman}}]{oliveira2008}%
  \BibitemOpen
  \bibfield  {author} {\bibinfo {author} {\bibfnamefont {M.~M.}\ \bibnamefont
  {de~Oliveira}}, \bibinfo {author} {\bibfnamefont {S.~G.}\ \bibnamefont
  {Alves}}, \bibinfo {author} {\bibfnamefont {S.~C.}\ \bibnamefont {Ferreira}},
  \ and\ \bibinfo {author} {\bibfnamefont {R.}~\bibnamefont {Dickman}},\ }\href
  {\doibase 10.1103/PhysRevE.78.031133} {\bibfield  {journal} {\bibinfo
  {journal} {Phys. Rev. E}\ }\textbf {\bibinfo {volume} {78}},\ \bibinfo
  {pages} {031133} (\bibinfo {year} {2008})}\BibitemShut {NoStop}%
\bibitem [{\citenamefont {de~Oliveira}\ \emph {et~al.}(2016)\citenamefont
  {de~Oliveira}, \citenamefont {Alves},\ and\ \citenamefont
  {Ferreira}}]{oliveira2016}%
  \BibitemOpen
  \bibfield  {author} {\bibinfo {author} {\bibfnamefont {M.~M.}\ \bibnamefont
  {de~Oliveira}}, \bibinfo {author} {\bibfnamefont {S.~G.}\ \bibnamefont
  {Alves}}, \ and\ \bibinfo {author} {\bibfnamefont {S.~C.}\ \bibnamefont
  {Ferreira}},\ }\href {\doibase 10.1103/PhysRevE.93.012110} {\bibfield
  {journal} {\bibinfo  {journal} {Phys. Rev. E}\ }\textbf {\bibinfo {volume}
  {93}},\ \bibinfo {pages} {012110} (\bibinfo {year} {2016})}\BibitemShut
  {NoStop}%
\bibitem [{\citenamefont {Harris}(1974)}]{harris1974}%
  \BibitemOpen
  \bibfield  {author} {\bibinfo {author} {\bibfnamefont {A.~B.}\ \bibnamefont
  {Harris}},\ }\href@noop {} {\bibfield  {journal} {\bibinfo  {journal} {J.
  Phys. C: Solid State Phys}\ }\textbf {\bibinfo {volume} {7}},\ \bibinfo
  {pages} {1671} (\bibinfo {year} {1974})}\BibitemShut {NoStop}%
\bibitem [{\citenamefont {Harris}(2016)}]{harris2016}%
  \BibitemOpen
  \bibfield  {author} {\bibinfo {author} {\bibfnamefont {A.~B.}\ \bibnamefont
  {Harris}},\ }\href@noop {} {\bibfield  {journal} {\bibinfo  {journal} {J.
  Phys.: Condens. Matter}\ }\textbf {\bibinfo {volume} {28}},\ \bibinfo {pages}
  {421006} (\bibinfo {year} {2016})}\BibitemShut {NoStop}%
\bibitem [{\citenamefont {Barghathi}\ and\ \citenamefont
  {Vojta}(2014)}]{barghathi2014}%
  \BibitemOpen
  \bibfield  {author} {\bibinfo {author} {\bibfnamefont {H.}~\bibnamefont
  {Barghathi}}\ and\ \bibinfo {author} {\bibfnamefont {T.}~\bibnamefont
  {Vojta}},\ }\href@noop {} {\bibfield  {journal} {\bibinfo  {journal} {Phys.
  Rev. Lett.}\ }\textbf {\bibinfo {volume} {113}},\ \bibinfo {pages} {120602}
  (\bibinfo {year} {2014})}\BibitemShut {NoStop}%
\bibitem [{\citenamefont {Schrauth}\ \emph
  {et~al.}(2018{\natexlab{a}})\citenamefont {Schrauth}, \citenamefont
  {Portela},\ and\ \citenamefont {Goth}}]{schrauth2018b}%
  \BibitemOpen
  \bibfield  {author} {\bibinfo {author} {\bibfnamefont {M.}~\bibnamefont
  {Schrauth}}, \bibinfo {author} {\bibfnamefont {J.~S.~E.}\ \bibnamefont
  {Portela}}, \ and\ \bibinfo {author} {\bibfnamefont {F.}~\bibnamefont
  {Goth}},\ }\href {\doibase 10.1103/PhysRevLett.121.100601} {\bibfield
  {journal} {\bibinfo  {journal} {Phys. Rev. Lett.}\ }\textbf {\bibinfo
  {volume} {121}},\ \bibinfo {pages} {100601} (\bibinfo {year}
  {2018}{\natexlab{a}})}\BibitemShut {NoStop}%
\bibitem [{\citenamefont {Schrauth}\ \emph
  {et~al.}(2018{\natexlab{b}})\citenamefont {Schrauth}, \citenamefont
  {Richter},\ and\ \citenamefont {Portela}}]{schrauth2018a}%
  \BibitemOpen
  \bibfield  {author} {\bibinfo {author} {\bibfnamefont {M.}~\bibnamefont
  {Schrauth}}, \bibinfo {author} {\bibfnamefont {J.~A.~J.}\ \bibnamefont
  {Richter}}, \ and\ \bibinfo {author} {\bibfnamefont {J.~S.~E.}\ \bibnamefont
  {Portela}},\ }\href@noop {} {\bibfield  {journal} {\bibinfo  {journal} {Phys.
  Rev. E}\ }\textbf {\bibinfo {volume} {97}},\ \bibinfo {pages} {022144}
  (\bibinfo {year} {2018}{\natexlab{b}})}\BibitemShut {NoStop}%
\bibitem [{\citenamefont {Cardy}(2012)}]{cardy2012}%
  \BibitemOpen
  \bibfield  {author} {\bibinfo {author} {\bibfnamefont {J.}~\bibnamefont
  {Cardy}},\ }\href@noop {} {\emph {\bibinfo {title} {{F}inite-{S}ize
  {S}caling}}},\ Vol.~\bibinfo {volume} {2}\ (\bibinfo  {publisher}
  {Elsevier},\ \bibinfo {year} {2012})\BibitemShut {NoStop}%
\bibitem [{\citenamefont {Kirkpatrick}\ \emph {et~al.}(1983)\citenamefont
  {Kirkpatrick}, \citenamefont {Gelatt},\ and\ \citenamefont
  {Vecchi}}]{kirkpatrick1983}%
  \BibitemOpen
  \bibfield  {author} {\bibinfo {author} {\bibfnamefont {S.}~\bibnamefont
  {Kirkpatrick}}, \bibinfo {author} {\bibfnamefont {C.~D.}\ \bibnamefont
  {Gelatt}}, \ and\ \bibinfo {author} {\bibfnamefont {M.~P.}\ \bibnamefont
  {Vecchi}},\ }\href@noop {} {\bibfield  {journal} {\bibinfo  {journal}
  {Science}\ }\textbf {\bibinfo {volume} {220}},\ \bibinfo {pages} {671}
  (\bibinfo {year} {1983})}\BibitemShut {NoStop}%
\bibitem [{\citenamefont {Bentley}(1975)}]{bentley1975}%
  \BibitemOpen
  \bibfield  {author} {\bibinfo {author} {\bibfnamefont {J.~L.}\ \bibnamefont
  {Bentley}},\ }\href {\doibase 10.1145/361002.361007} {\bibfield  {journal}
  {\bibinfo  {journal} {Commun. ACM}\ }\textbf {\bibinfo {volume} {18}},\
  \bibinfo {pages} {509} (\bibinfo {year} {1975})}\BibitemShut {NoStop}%
\bibitem [{Note1()}]{Note1}%
  \BibitemOpen
  \bibinfo {note} {That we can always assign an even number $q_i$ of neighbors
  to any set of sites can be easily seen by arranging the sites in a circle and
  connecting every site to its $q_i/2$ left (or right) neighbors.}\BibitemShut
  {Stop}%
\bibitem [{\citenamefont {Euler}(1741)}]{euler1741}%
  \BibitemOpen
  \bibfield  {author} {\bibinfo {author} {\bibfnamefont {L.}~\bibnamefont
  {Euler}},\ }\href@noop {} {\bibfield  {journal} {\bibinfo  {journal}
  {Commentarii academiae scientiarum Petropolitanae}\ ,\ \bibinfo {pages}
  {128}} (\bibinfo {year} {1741})}\BibitemShut {NoStop}%
\bibitem [{\citenamefont {Su}\ and\ \citenamefont {Drysdale}(1997)}]{su1997}%
  \BibitemOpen
  \bibfield  {author} {\bibinfo {author} {\bibfnamefont {P.}~\bibnamefont
  {Su}}\ and\ \bibinfo {author} {\bibfnamefont {R.~L.~S.}\ \bibnamefont
  {Drysdale}},\ }\href@noop {} {\bibfield  {journal} {\bibinfo  {journal}
  {Comput. Geom.}\ }\textbf {\bibinfo {volume} {7}},\ \bibinfo {pages} {361}
  (\bibinfo {year} {1997})}\BibitemShut {NoStop}%
\bibitem [{\citenamefont {Supowit}(1983)}]{supowit1983}%
  \BibitemOpen
  \bibfield  {author} {\bibinfo {author} {\bibfnamefont {K.~J.}\ \bibnamefont
  {Supowit}},\ }\href@noop {} {\bibfield  {journal} {\bibinfo  {journal}
  {J.~ACM}\ }\textbf {\bibinfo {volume} {30}},\ \bibinfo {pages} {428}
  (\bibinfo {year} {1983})}\BibitemShut {NoStop}%
\bibitem [{\citenamefont {Vaidya}(1989)}]{vaidya1989}%
  \BibitemOpen
  \bibfield  {author} {\bibinfo {author} {\bibfnamefont {P.~M.}\ \bibnamefont
  {Vaidya}},\ }\href {\doibase 10.1007/BF02187718} {\bibfield  {journal}
  {\bibinfo  {journal} {Discrete Comput. Geom.}\ }\textbf {\bibinfo {volume}
  {4}},\ \bibinfo {pages} {101} (\bibinfo {year} {1989})}\BibitemShut {NoStop}%
\bibitem [{\citenamefont {Dwyer}(1987)}]{dwyer1987}%
  \BibitemOpen
  \bibfield  {author} {\bibinfo {author} {\bibfnamefont {R.~A.}\ \bibnamefont
  {Dwyer}},\ }\href@noop {} {\bibfield  {journal} {\bibinfo  {journal}
  {Algorithmica}\ }\textbf {\bibinfo {volume} {2}},\ \bibinfo {pages} {137}
  (\bibinfo {year} {1987})}\BibitemShut {NoStop}%
\bibitem [{\citenamefont {Agarwal}\ and\ \citenamefont
  {Matau{\v{s}}ek}(1992)}]{agarwal1992}%
  \BibitemOpen
  \bibfield  {author} {\bibinfo {author} {\bibfnamefont {P.~K.}\ \bibnamefont
  {Agarwal}}\ and\ \bibinfo {author} {\bibfnamefont {J.}~\bibnamefont
  {Matau{\v{s}}ek}},\ }in\ \href@noop {} {\emph {\bibinfo {booktitle}
  {Proceedings of the third annual ACM-SIAM symposium on Discrete
  algorithms}}}\ (\bibinfo {organization} {Society for Industrial and Applied
  Mathematics},\ \bibinfo {year} {1992})\ pp.\ \bibinfo {pages}
  {58--65}\BibitemShut {NoStop}%
\bibitem [{\citenamefont {Dunham}(1986)}]{dunham1986}%
  \BibitemOpen
  \bibfield  {author} {\bibinfo {author} {\bibfnamefont {D.}~\bibnamefont
  {Dunham}},\ }in\ \href {\doibase
  https://doi.org/10.1016/B978-0-08-033986-3.50018-5} {\emph {\bibinfo
  {booktitle} {Symmetry}}}\ (\bibinfo  {publisher} {Pergamon},\ \bibinfo {year}
  {1986})\ pp.\ \bibinfo {pages} {139 -- 153}\BibitemShut {NoStop}%
\bibitem [{\citenamefont {Sausset}\ and\ \citenamefont
  {Tarjus}(2007)}]{sausset2007}%
  \BibitemOpen
  \bibfield  {author} {\bibinfo {author} {\bibfnamefont {F.}~\bibnamefont
  {Sausset}}\ and\ \bibinfo {author} {\bibfnamefont {G.}~\bibnamefont
  {Tarjus}},\ }\href@noop {} {\bibfield  {journal} {\bibinfo  {journal} {J.
  Phys. A: Math. Theor.}\ }\textbf {\bibinfo {volume} {40}},\ \bibinfo {pages}
  {12873} (\bibinfo {year} {2007})}\BibitemShut {NoStop}%
\bibitem [{\citenamefont {Schrauth}\ \emph {et~al.}()\citenamefont {Schrauth},
  \citenamefont {Goth},\ and\ \citenamefont {D{\"o}ring}}]{marqov}%
  \BibitemOpen
  \bibfield  {author} {\bibinfo {author} {\bibfnamefont {M.}~\bibnamefont
  {Schrauth}}, \bibinfo {author} {\bibfnamefont {F.}~\bibnamefont {Goth}}, \
  and\ \bibinfo {author} {\bibfnamefont {M.}~\bibnamefont {D{\"o}ring}},\
  }\href@noop {} {\bibinfo  {journal} {{to be published}}\ }\BibitemShut
  {NoStop}%
\bibitem [{\citenamefont {Nattermann}(1998)}]{nattermann1998}%
  \BibitemOpen
\bibfield  {journal} {  }\bibfield  {author} {\bibinfo {author} {\bibfnamefont
  {T.}~\bibnamefont {Nattermann}},\ }in\ \href@noop {} {\emph {\bibinfo
  {booktitle} {{S}pin {G}lasses and {R}andom {F}ields}}}\ (\bibinfo
  {publisher} {World Scientific},\ \bibinfo {year} {1998})\ pp.\ \bibinfo
  {pages} {277--298}\BibitemShut {NoStop}%
\bibitem [{\citenamefont {Binder}\ and\ \citenamefont
  {Young}(1986)}]{binder1986}%
  \BibitemOpen
  \bibfield  {author} {\bibinfo {author} {\bibfnamefont {K.}~\bibnamefont
  {Binder}}\ and\ \bibinfo {author} {\bibfnamefont {A.~P.}\ \bibnamefont
  {Young}},\ }\href@noop {} {\bibfield  {journal} {\bibinfo  {journal} {\rmp}\
  }\textbf {\bibinfo {volume} {58}},\ \bibinfo {pages} {801} (\bibinfo {year}
  {1986})}\BibitemShut {NoStop}%
\bibitem [{\citenamefont {Wolff}(1989)}]{wolff1989}%
  \BibitemOpen
  \bibfield  {author} {\bibinfo {author} {\bibfnamefont {U.}~\bibnamefont
  {Wolff}},\ }\href {\doibase 10.1103/PhysRevLett.62.361} {\bibfield  {journal}
  {\bibinfo  {journal} {Phys. Rev. Lett.}\ }\textbf {\bibinfo {volume} {62}},\
  \bibinfo {pages} {361} (\bibinfo {year} {1989})}\BibitemShut {NoStop}%
\bibitem [{\citenamefont {Metropolis}\ \emph {et~al.}(1953)\citenamefont
  {Metropolis}, \citenamefont {Rosenbluth}, \citenamefont {Rosenbluth},
  \citenamefont {Teller},\ and\ \citenamefont {Teller}}]{metropolis1953}%
  \BibitemOpen
  \bibfield  {author} {\bibinfo {author} {\bibfnamefont {N.}~\bibnamefont
  {Metropolis}}, \bibinfo {author} {\bibfnamefont {A.~W.}\ \bibnamefont
  {Rosenbluth}}, \bibinfo {author} {\bibfnamefont {M.~N.}\ \bibnamefont
  {Rosenbluth}}, \bibinfo {author} {\bibfnamefont {A.~H.}\ \bibnamefont
  {Teller}}, \ and\ \bibinfo {author} {\bibfnamefont {E.}~\bibnamefont
  {Teller}},\ }\href@noop {} {\bibfield  {journal} {\bibinfo  {journal} {J.
  Chem. Phys.}\ }\textbf {\bibinfo {volume} {21}},\ \bibinfo {pages} {1087}
  (\bibinfo {year} {1953})}\BibitemShut {NoStop}%
\bibitem [{\citenamefont {Hasenbusch}(2010)}]{hasenbusch2010}%
  \BibitemOpen
  \bibfield  {author} {\bibinfo {author} {\bibfnamefont {M.}~\bibnamefont
  {Hasenbusch}},\ }\href@noop {} {\bibfield  {journal} {\bibinfo  {journal}
  {Physical Review B}\ }\textbf {\bibinfo {volume} {82}},\ \bibinfo {pages}
  {174433} (\bibinfo {year} {2010})}\BibitemShut {NoStop}%
\bibitem [{\citenamefont {Selke}\ and\ \citenamefont
  {Shchur}(2005)}]{selke2005}%
  \BibitemOpen
  \bibfield  {author} {\bibinfo {author} {\bibfnamefont {W.}~\bibnamefont
  {Selke}}\ and\ \bibinfo {author} {\bibfnamefont {L.}~\bibnamefont {Shchur}},\
  }\href@noop {} {\bibfield  {journal} {\bibinfo  {journal} {J. Phys. A: Math.
  Gen.}\ }\textbf {\bibinfo {volume} {38}},\ \bibinfo {pages} {L739} (\bibinfo
  {year} {2005})}\BibitemShut {NoStop}%
\bibitem [{\citenamefont {Selke}\ and\ \citenamefont
  {Shchur}(2009)}]{selke2009}%
  \BibitemOpen
  \bibfield  {author} {\bibinfo {author} {\bibfnamefont {W.}~\bibnamefont
  {Selke}}\ and\ \bibinfo {author} {\bibfnamefont {L.}~\bibnamefont {Shchur}},\
  }\href@noop {} {\bibfield  {journal} {\bibinfo  {journal} {Phys. Rev. E}\
  }\textbf {\bibinfo {volume} {80}},\ \bibinfo {pages} {042104} (\bibinfo
  {year} {2009})}\BibitemShut {NoStop}%
\bibitem [{\citenamefont {Malakis}\ \emph {et~al.}(2014)\citenamefont
  {Malakis}, \citenamefont {Fytas},\ and\ \citenamefont
  {G{\"u}lpinar}}]{malakis2014}%
  \BibitemOpen
  \bibfield  {author} {\bibinfo {author} {\bibfnamefont {A.}~\bibnamefont
  {Malakis}}, \bibinfo {author} {\bibfnamefont {N.~G.}\ \bibnamefont {Fytas}},
  \ and\ \bibinfo {author} {\bibfnamefont {G.}~\bibnamefont {G{\"u}lpinar}},\
  }\href@noop {} {\bibfield  {journal} {\bibinfo  {journal} {Physical Review
  E}\ }\textbf {\bibinfo {volume} {89}},\ \bibinfo {pages} {042103} (\bibinfo
  {year} {2014})}\BibitemShut {NoStop}%
\bibitem [{\citenamefont {Ferrenberg}\ \emph {et~al.}(2018)\citenamefont
  {Ferrenberg}, \citenamefont {Xu},\ and\ \citenamefont
  {Landau}}]{ferrenberg2018}%
  \BibitemOpen
  \bibfield  {author} {\bibinfo {author} {\bibfnamefont {A.~M.}\ \bibnamefont
  {Ferrenberg}}, \bibinfo {author} {\bibfnamefont {J.}~\bibnamefont {Xu}}, \
  and\ \bibinfo {author} {\bibfnamefont {D.~P.}\ \bibnamefont {Landau}},\
  }\href {\doibase 10.1103/PhysRevE.97.043301} {\bibfield  {journal} {\bibinfo
  {journal} {Phys. Rev. E}\ }\textbf {\bibinfo {volume} {97}},\ \bibinfo
  {pages} {043301} (\bibinfo {year} {2018})}\BibitemShut {NoStop}%
\bibitem [{\citenamefont {Keblinski}\ \emph {et~al.}(2002)\citenamefont
  {Keblinski}, \citenamefont {Bazant}, \citenamefont {Dash},\ and\
  \citenamefont {Treacy}}]{keblinski2002}%
  \BibitemOpen
  \bibfield  {author} {\bibinfo {author} {\bibfnamefont {P.}~\bibnamefont
  {Keblinski}}, \bibinfo {author} {\bibfnamefont {M.}~\bibnamefont {Bazant}},
  \bibinfo {author} {\bibfnamefont {R.}~\bibnamefont {Dash}}, \ and\ \bibinfo
  {author} {\bibfnamefont {M.}~\bibnamefont {Treacy}},\ }\href@noop {}
  {\bibfield  {journal} {\bibinfo  {journal} {Physical Review B}\ }\textbf
  {\bibinfo {volume} {66}},\ \bibinfo {pages} {064104} (\bibinfo {year}
  {2002})}\BibitemShut {NoStop}%
\bibitem [{\citenamefont {B{\"u}chner}\ and\ \citenamefont
  {Heyde}(2017)}]{buchner2017}%
  \BibitemOpen
  \bibfield  {author} {\bibinfo {author} {\bibfnamefont {C.}~\bibnamefont
  {B{\"u}chner}}\ and\ \bibinfo {author} {\bibfnamefont {M.}~\bibnamefont
  {Heyde}},\ }\href@noop {} {\bibfield  {journal} {\bibinfo  {journal} {Prog.
  Surf. Sci.}\ }\textbf {\bibinfo {volume} {92}},\ \bibinfo {pages} {341}
  (\bibinfo {year} {2017})}\BibitemShut {NoStop}%
\bibitem [{\citenamefont {Gaspard}(1985)}]{gaspard1985}%
  \BibitemOpen
  \bibfield  {author} {\bibinfo {author} {\bibfnamefont {J.}~\bibnamefont
  {Gaspard}},\ }\href@noop {} {\bibfield  {journal} {\bibinfo  {journal} {J.
  Phys. Colloq.}\ }\textbf {\bibinfo {volume} {46}},\ \bibinfo {pages} {C8}
  (\bibinfo {year} {1985})}\BibitemShut {NoStop}%
\end{thebibliography}
%

\end{document}